\documentclass{elsarticle}
\pdfoutput=1
\usepackage{graphicx, color}
\usepackage{tikz}
\usepackage{pgfplots}

\newcommand{\pder}[2][]{\frac{\partial#1}{\partial#2}}

\begin{document}
\begin{frontmatter}
\title{Nonlinear pressure-velocity waveforms  in large arteries, shock waves and wave separation }

\author{Oleg  Ilyin}

\address{
Dorodnicyn Computing Center of Federal Research Center “Computer Science and Control” of RAS,  Vavilova  40, Moscow, 119333, Russia
}
\ead{oilyin@gmail.com}

\begin{abstract}
Nonlinear inviscid 1D blood flow equations were studied analytically using  the method of characteristics.
The  initial-boundary value  problem for a class  of the initial-boundary conditions (including a triangle-shaped profile)  at the aortic outlet was  considered. The nonlinear  pressure-velocity profile and  shock conditions were  derived as closed analytical expressions with  second order  accuracy on the relative luminal area change. Finally, the fully nonlinear wave separation expressions were obtained using the Riemann invariants. The   results  show  good  correspondence  with the data from the literature.
\end{abstract}


\end{frontmatter}


\section{Introduction}
The  most popular 1D   hemodynamic pressure-velocity propagation models  are based  on the  quasilinear partial differential equations  of the hyperbolic type \cite{1973hughes} -\cite{2003sherwin}.
The equations describe   blood motion  in  distensible vessels in terms of  unknown pressure and  blood velocity (or the vessel's
cross -sectional area and   blood  velocity).
Solutions to 1D equations are usually obtained  numerically for the branching arterial network \cite{1992stergioupulos}-\cite{2011alastruey} and they  can  be  coupled with the solutions of the full 3D Navier-Stokes equations
\cite{2001formaggia}.
 As the difference $\Delta A$ between  the  systolic luminal area $A_1$ and the diastolic $A_0$  is small  ($\Delta A/A_0$ is assumed  to be  negligible) the the equations  can be  linearized around $A_0$. The reduced equations are well known in sound  propagation theory \cite{1945lamb}. The  linear theory allows the assessment of several properties of the flow which are important for the estimation of cardiovascular risk. Any change  in the geometry of a vessel (bifurcation, lumen narrowing and other factors changing vascular impedance)  causes  an appearance of reflected waves. Simple and  concise formulas for the wave separation into the  forward and    backward waves can be  obtained  in the  framework of the linear theory using the wave separation analysis \cite{1972westerhof}-\cite{2015westerhof}
or  wave intensity analysis \cite{2009hughes}-\cite{2001khir}.

The perturbations travel along the constant and  parallel characteristics in the linear approximation and  hence the waves do  not  change  their initial form and shock-waves do not appear. Nevertheless, the nonlinear effects  can play a significant  role in the behaviour of the pressure-velocity waves.
For instance, if the reported difference  between the diameters of large vessels like the carotid artery  in  systole and   diastole is $10\%$ \cite{1994gamble},  then the relative difference  between the lumen areas are  about  $\Delta A/A_0 \approx 0.2$. Therefore, a full nonlinear analysis should  be performed.
The existence of  shock-waves has been proven numerically in \cite{1981forbes}, while
the formation of  shock waves at a distances between  few centimetres   and  several meters from the aortic outlet  (depending  on the elastic properties)  has  been reported in  \cite{1970rudinger}. The possibility of roll waves  in collapsible tubes has  been considered in \cite{1999brook}.
The effects of  friction coupled with nonlinearities  has  been studied in \cite{1995pythoud}-\cite{1996pythoud}.

The goal of the  paper is to derive  the closed analytical expression for the blood pressure, the luminal area and the velocity using the nonlinear equations for the mass  and moment transfer.
The flow  is considered to be inviscid, while the viscoelastic terms in the wall-pressure  response  are excluded and  the  unperturbed luminal area is constant within  the vessel.
In contrast to the linear theory, where $\Delta A/A_0$ is assumed  to be negligible, we will include these  terms but exclude quadratic terms $(\Delta A/A_0)^2$. In other words, the  obtained solutions include $(\Delta A/A_0)^2$ error in  comparison to the exact solutions of the mass  and  moment  transfer equations.
The simplified nonlinear assumption allows the main features of the  nonlinear theory to be included  as  the characteristics can converge or  diverge.
The  analytical expression for the pressure-velocity waves for a class  of the initial-boundary conditions is derived   providing  quantitative  information about  the evolution of  waves in time and space.  It is shown that the wave significantly changes  its shape  while traveling  along  a vessel and  shock waves can  emerge.
Moreover,  with the use  of the Riemann invariants  the  closed nonlinear wave separation formulae were calculated, the  analytical expressions for the forward and  backward waves were obtained.

\section{The mass  and  momentum conservation equations and their reduction in the case  of sole wave}

Consider the mass and  momentum conservation equations for inviscid blood motion in a distensible  vessel with
 impermeable walls   \cite{2003sherwin}
 \begin{equation}\label{04onedeq2}
\pder[A]{t}+\pder[Q]{x}=0, \quad \pder[Q]{t}+\frac{\partial}{\partial x} \left(\frac{\alpha Q^2}{A} \right)=-\frac{A}{\rho}\pder[p]{x},
\end{equation}
where $Q,A$ are the flow, the cross-section area of a vessel respectively, $\alpha$  is the profile shape factor assumed to be equal
unity (the flow  is  flat) and
$\rho$  is the blood density.

We introduce the  distensibility $D(A)$  which is by the definition  \cite{2003sherwin}
\begin{equation}\label{03dist}
D(A)=\frac{1}{A}\frac{dA}{dp}.
\end{equation}
We assume  that  distensibility $D(A)$ is a known function of $A$. In most applications this  relationship is as follows
$$
D(A)=\frac{D_0}{A^{n}}, \quad  n>0.
$$
Integrating the expression (\ref{03dist}) we obtain
\begin{equation}\label{p3exp_p}
p=p_0+\frac{1}{n D_0}(A^{n}-A_0^{n}),
\end{equation}
where $A_0$ is  the  undisturbed luminal area  $p_0$ is the corresponding blood  pressure. Additional information about
pressure-area relationship and  the Hooke law could be found  in the Appendix.

For a flat velocity  profile  ($\alpha=1$) we use $Q=uA$. In terms of $u,A$ the equations (\ref{04onedeq2}) have the form
 \begin{equation}\label{05onedeq}
\pder[A]{t}+u\pder[A]{x}+A\pder[u]{x}=0, \quad \pder[u]{t}+\frac{1}{\rho D(A)A}\pder[A]{x}+u\pder[u]{x}=0.
\end{equation}
For a sole forward traveling wave we  have  the  following  relationship between the area of the vessel and  the  velocity (see paragraph 4)
\begin{equation}\label{06subst}
u=\int_{A_0}^{A}\frac{dz}{\sqrt{\rho D(z)}z},
\end{equation}
where $A_0$ is the undisturbed vessel cross-sectional area
 (for the  backward  traveling  wave we substitute $u=-\int_{A_0}^{A}\frac{dz}{\sqrt{\rho D(z)}z}$).
If we substitute the expression (\ref{06subst}) in the system (\ref{05onedeq}) then  the both equations in (\ref{05onedeq}) will have the same form, hence  the system of the partial  differential equations reduce  to   only one  equation
$$
\pder[A]{t}+\left (\int_{A_0}^{A}\frac{dz}{\sqrt{\rho D(z)}z}+\frac{1}{\sqrt{\rho D(A)}}  \right )\pder[A]{x}=0.
$$
Taking  into account that $D(A)=D_0/A^n$ we finally  obtain the equation for  the forward wave
\begin{equation}\label{05onedeq1}
\pder[A]{t}+\frac{1}{\sqrt{\rho D_0}}\left (\left(1+\frac{2}{n}\right)A^{n/2}-\frac{2}{n}A_0^{n/2}  \right )\pder[A]{x}=0.
\end{equation}
The analytical solution to this equation in an implicit form is  presented  in Appendix 2.

\section{Analytical nonlinear area-pressure wave and shock wave conditions}

The equation (\ref{05onedeq1}) can be reduced to the following  (see details below)
\begin{equation}\label{05beq}
\frac{\partial A}{\partial t}+c_0\left(1+\frac{n+1}{2A_0}(A-A_0)\right)\frac{\partial A}{\partial x}=0,
\end{equation}
 where $c_0=\sqrt{A_0^n/\rho D_0}$ is the  pulse wave velocity  in the  linear theory. We  supplement this equation  with the following triangle-shaped initial-boundary  condition (Fig. 2)  at $x=0$ for $t \in [0,T_0]$, where $T_0$ is  one  heartbeat
\begin{equation}\label{05bounds1}
A(t,x)|_{x=0}=A_0+at,\quad t\in [0, t_0),
\end{equation}
\begin{equation}\label{05bounds2}
A(t,x)|_{x=0}=A_0+b(t_1-t), \quad  b\equiv \frac{at_0}{t_1-t_0} ,\quad t \in [t_0, t_1],
\end{equation}
\begin{equation}\label{05bounds3}
A(t,x)|_{x=0}=A_0,\quad t \in (t_1,T_0]
\end{equation}
and
\begin{equation}\label{05bounds4}
A(t,x)|_{t=0}=A_0,\quad x>0.
\end{equation}
In practice $T_0 \approx 1 s, t_1 \approx 0.3 s$.
The  conditions (\ref{05bounds1})-(\ref{05bounds2}) correspond  to systole and (\ref{05bounds3})  to diastole.
We do not consider  the exponential drop of the cross-sectional area or  pressure during diastole which is usually observed  in clinical studies. This effect  is a consequence  of the resistance of  the  distal vasculature which is not included in the  present investigation and  the  cross-sectional area  is assumed  to be constant.

The nonlinear  study of  blood  motion equations with the use of the method of characteristics has been presented in several papers \cite{1990parker}-\cite{lbv}.
The  analytical expression for the  pressure-velocity (or the area-velocity) waves has  not  been presented before in the nonlinear case; our  goal is to solve analytically the problem  (\ref{05beq}) and (\ref{05bounds1})-(\ref{05bounds4}).

From Eq.  (\ref{05onedeq1}) one can deduce that  the nonlinear area-pressure pulse velocity $c$  has the following form
$$
c=\frac{1}{\sqrt{\rho D_0}}\left (\left(1+\frac{2}{n}\right)A^{n/2}-\frac{2}{n}A_0^{n/2}  \right ),
$$
then the relative velocity (analog of the Mach number) is
$$
\frac{u}{c}=\frac{2}{n}\frac{(A^{n/2}-A_0^{n/2})}{\left(1+\frac{2}{n}\right)A^{n/2}-\frac{2}{n}A_0^{n/2} },
$$
where we have applied the expression (\ref{06subst})  for the blood velocity $u=\int_{A_0}^{A}\frac{ds}{\sqrt{\rho D(s)}s}=\frac{2(A^{n/2}-A_0^{n/2})}{n\sqrt{\rho D_0}}$.
If one assumes that the  value of the perturbed cross-sectional area $A$  is $20\%$ larger than the unperturbed area $A_0$ i.e. $A/A_0=1.2$ and $n=0.5$ then the following result is obtained
$$
\frac{u}{c} \approx 0.218,
$$
therefore, similarly to the clinical measurements the flow is significantly subsonic.

The relative change of the cross-sectional is  not small then this effect should be  kept in the consideration.  We
expand the  pulse velocity  in Taylor series near $A_0$,  keep only the  linear perturbations and  neglect quadratic  ($(A/A_0)^2 \approx 1.04 $) and  higher terms
\begin{equation}\label{05pulsevel}
c \approx c_0\left(1+\frac{n+1}{2A_0}(A-A_0) \right),
\end{equation}
where $c_0=\sqrt{A_0^n/\rho D_0}$ is the  pulse wave velocity  in the  linear theory. Replacing  the pulse wave velocity in Eq. (\ref{05onedeq1})
by its value from (\ref{05pulsevel}) we derive the equation (\ref{05beq}).  The solution to Eq. (\ref{05beq})  can be constructed in the same way as for
Eq. (\ref{05onedeq1}) (Appendix 2), we have
$$
A-A_0=a\left(t-\frac{x}{c_0(1+\frac{n+2}{2A_0}(A-A_0))} \right),
$$
where $c_0=\sqrt{A_0^{n}/\rho D_0}$ is the pulse  wave velocity in the linear theory.
Solving the quadratic algebraic equation for $A-A_0$  we  finally deduce  that
 \begin{equation}\label{07sol1}
A=A_0+\frac{-(1-akt)+\sqrt{(1-akt)^2-4ak(c_0^{-1}x-t)}}{2k},
\end{equation}
where $k\equiv\frac{n+2}{2A_0}$. The expression (\ref{07sol1}) gives the solution of the considered problem
for the boundary condition (\ref{05bounds1}).

Similarly, the boundary condition  (\ref{05bounds2}) leads to the  solution
$$
A=A_0+
$$
\begin{equation}\label{07sol2}
+\frac{-(1-bk(t_1-t))+\sqrt{(1-bk(t_1-t))^2+4bk(c_0^{-1}x+(t_1-t))}}{2k}.
\end{equation}


Now we need  to obtain the  conditions for the appearance of a shock wave. The shock wave appears when $\frac{\partial A}{\partial t}, \frac{\partial A}{\partial x}$ are infinite. This  happens  when  the denominator in $\frac{\partial A}{\partial t}, \frac{\partial A}{\partial x}$ equals  to zero.
Then for  (\ref{07sol1}) we have the curve $C_{s}$
$$
C_s:\quad  x=\frac{c_0}{4ak}(1+akt)^2.  
$$
We conclude  that the shock wave appears for the first time $t_s$ when the characteristic $x=c_0t$ crosses $C_s$. Then
$$
1+akt_s=\sqrt{4akt_s},
$$
the  solution of this  algebraic equation is
$$
t_s=\frac{1}{ka}
$$
and finally
\begin{equation}\label{08shock}
x_s=\frac{c_0}{ka}=\frac{2c_0t_0}{(n+2)\frac{at_0}{A_0}}= \frac{2c_0t_0}{(n+2)}\left(\frac{\Delta A}{A_0} \right)^{-1},
\end{equation}
where  $\Delta A=a_0t_0$.
If we assume  that $c_0$  lies  in the range from $3\, m/s$  to $10\, m/s$ and $n=0.5, at_0\equiv \Delta A=0.2A_0$ where $t_0=0.15 s$,
then  we  have for the distance of shock $x_s$  the estimate $4c_0t_0$  varying from  $1.8\, m$  to $6\, m$. This result is similar to the estimate from \cite{1970rudinger}.

The detailed consideration of the wave dynamics after the  appearance  of the shock requires the full analysis of the characteristics and is presented below.
For the sake  of simplicity  we set $a=b$ for the  boundary conditions (symmetric initial pulse) (\ref{05bounds1})-(\ref{05bounds2}), we have
$A(t,x)|_{x=0}=A_0+at,\, t\in [0, t_0)$ and $A(t,x)|_{x=0}=A_0+a(2t_0-t) ,\, t \in [t_0, 2t_0]$.
The  solution along  the characteristics  is given by
\begin{equation}\label{05beq_sol}
.A=f(s),  \quad t(s)=\frac{1}{c_0\bigl(1+k(f(s)-A_0)\bigr)}x+s,
\end{equation}
where $s$ is a parameter (constant of integration),  the function $f(s)$ is  unknown, we can  find $f(s)$  from the boundary condition.
From the conditions at $x=0$  we deduce  that $f(s)$ equals $A_0+as, s\in [0, t_0)$ and $A_0+a(2t_0-s) ,\, s \in [t_0, 2t_0]$.
The shock wave  $X(t)$ appears if two characteristics  (fixed by the parameters $s=s_1, s=s_2$) intersect. Using (\ref{05beq_sol}) we have
\begin{equation}\label{05beq_char}
X=c_0(1+k(f(s_1)-A_0))(t-s_1),  \quad X=c_0(1+k(f(s_2)-A_0))(t-s_2).
\end{equation}
We  need  to find  three unknowns $s_1, s_2, X$, but  only the two equations are introduced. The  conditions above  define all possible intersections between the characteristic lines, but  we need to keep only a subset of the intersections  which define a shock curve.  Additional information about the shock wave velocity can be  extracted from the equation (\ref{05beq}). Integrating  Eq. (\ref{05beq}) on $x$  we deduce  that
$$
\int_0^{\infty} (A(t,x)-A_0) dx
$$
is  conserving quantity (integral of motion) for  $t>t_1$. This fact will be  important in the further consideration of  large $t$ pulse wave asymptotics.

Eq. (\ref{05beq})  can be rewritten in the following form consistent with the aforementioned  integral of motion
$$
\frac{\partial A}{\partial t}+\frac{\partial Q}{\partial x}=0,  \quad Q \equiv c_0(A-A_0)+\frac{c_0k}{2}(A-A_0)^2.
$$
Integration of this equation  over $x$  leads to the Hugoniot jump condition
$$
\frac{dX}{dt}=\frac{Q_1-Q_2}{A_1-A_2},
$$
where $Q_1, Q_2$ and $A_1, A_2$ are  the values of the corresponding variables in the right and the left sides from the jump. Then the shock velocity equals
$$
\frac{dX}{dt}=
$$
$$
=\frac{c_0(A(s_1)-A_0)+\frac{c_0k}{2}(A(s_1)-A_0)^2-(c_0(A(s_2)-A_0)+\frac{c_0k}{2}(A(s_2)-A_0)^2)}{(A(s_1)-A_0)-(A(s_2)-A_0)}
$$
and we  know that $A(s_1)=f(s_1), A(s_2)=f(s_2)$ ,   finally
\begin{equation}\label{05beq_shockveloc}
\frac{dX}{dt}=c_0+\frac{c_0k}{2}\bigl\{f(s_1)+f(s_2)-2A_0\bigr\}.
\end{equation}
Eqs  (\ref{05beq_char}) and (\ref{05beq_shockveloc}) allow to  express $s_1, s_2$ via $X$ and  derive  the closed differential equation for $X(t)$. The form of this equation depends  on the selection of $s_1, s_2$.  If $s_1, s_2 \in [0, t_0)$ then from (\ref{05beq_char}) we obtain
$$
s_{1,2}=\frac{(akt-1) \pm \sqrt{(1+akt)^2-\frac{4ak}{c_0}X}}{2ak}
$$
and  Eq. (\ref{05beq_shockveloc}) takes the form
$$
\frac{dX}{dt}=c_0+\frac{c_0k}{2} (akt-1),
$$
which has the solution
$ X(t)=\left(x_0-\frac{c_0}{4ak}\right)+\frac{c_0}{4ak}(1+akt)^2, $
where $x_0$  is  constant of integration. The case $s_2=0$ corresponds  to the characteristic $x=c_0t$, this characteristic intersects
the shock $X(t)$ at the point  $(t_s, x_s)$ (Figure 1), then at this  point we have the additional constraint $X(t_s)=c_0t_s$ and we conclude that $x_0=\frac{c_0}{4ak}$. Therefore, as  anticipated,  we recover the formula
\begin{equation}\label{05beq_cs}
C_s:\quad X(t)=\frac{c_0}{4ak}(1+akt)^2,\quad t \in [t_s, t_s^*],
\end{equation}
 which was obtained  using  the explicit forms of the wave fronts. Note  that $t_s=\frac{1}{ak}$  and $t_s^*=\frac{1+2akt_0}{ak}$,
 where  $t_s^*$ denotes the moment  of time when the characteristic $x=c_0(1+kat_0)(t-t_0)$  meets the curve $C_s$ (Fig. 1).

The  remained  case  corresponds  to the shock wave  originated from the intersection of the characteristics  $s_1 \in [0, t_0),
s_2 \in [t_0, 2t_0] $. This case is  more complicated.
 Similarly to the  previous  case  we obtain  for the velocity of the shock wave the equation
$$
C_s:\quad \frac{dX}{dt}=c_0+\frac{c_0}{4a}\Bigl \{ 2ak(t_0+t)+\sqrt{(1+akt)^2+\frac{4ak}{c_0}X}-
$$
\begin{equation}\label{05beq_dom2vel}
  -\sqrt{(1+2akt_0-akt)^2+\frac{4ak}{c_0}X}  \Bigr \}, \quad t>t_s^*.
\end{equation}
 The equation (\ref{05beq_dom2vel})  can be  investigated qualitatively.
 We can notice that the term in the curved brackets is  positive for large  $t$. Therefore,  the distance  between the characteristic with $s=t_1=2t_0$ and  the  position of  the shock is monotonically increasing  since  the propagation  velocity of the  shock (\ref{05beq_dom2vel}) is  greater than the velocity of the considered characteristic which equals $c_0$.

 The asymptotic behavior  of the solutions  (\ref{05beq_dom2vel}) can be also obtained. If $t \rightarrow \infty$ then the difference between the square root terms tends to zero and then $X_t=c_0+\frac{c_0k}{2}(t_0+t)$, the solution behaves as $c_0(1+\frac{kt_0}{2})t+\frac{c_0k}{4}t^2$.

 This  fact means  that  the length of the  pulse wave tends  to infinity for large $t$ (infinitely stretched wave). Moreover, the maximum amplitude of the wave decreases and  tends to $A_0$  after appearance  of the shock because $\int_0^{\infty} (A-A_0) dx$ is conserving quantity.

\begin{figure}
\begin{tikzpicture}[>=latex]

 \draw [<->] [thick] (0,10) -- (0,0) -- (10,0);
 \node [below] at (10,0) {$x$};
 \node [left] at (0,10) {$t$};
 \node [below left] at (0,0) {$0$};

\draw [ultra thick] (0,0) -- (3, 1.69);
 \node [below] at (2,0.7) {$x=c_0 t$};

 \node [below] at (4.5, 2) {$x_s=\frac{c_0}{ak}, t_s=\frac{1}{ak}$} ;

 \draw [dashed] (0,3) -- (4.1, 3.4);
 \node [left] at (0,3) {$t_0$};
 \node [below] at (2, 4.1) {$x=c_0(1+kat_0)(t-t_0)$};
 \node [below] at (6.5 , 3.8) {$x_s^*=c_0\frac{(1+akt_0)^2}{ak} , t_s^*=\frac{1+2akt_0}{ak}$};

 \draw [ultra thick] (0,6) -- (6.6,10);
 \node [left] at (0,6) {$t_1$};
 \node [left] at (9,10) {$x=c_0(t-t_1)$};

 \draw [ultra thick] (3, 1.69) to [out=65,in=190] (8.5, 6);
 \node [left] at (7, 5) {$C_s$};


 \draw [thick] (1,1.5) rectangle (2,2.5);
 \node  at (1.5,2) {$\mathbf{1}$};

 \draw [thick] (1,5) rectangle (2,6);
 \node  at (1.5,5.5) {$\mathbf{2}$};

\end{tikzpicture}
\caption{ The characteristics for the problem (\ref{05onedeq1})-(\ref{05bounds4}). The domain $1$ is bounded by the characteristics $x=c_0t, x=c_0(1+kat_0)(t-t_0)$  and  the  shock curve $C_s$ , the  domain $2$ is bounded by the characteristics $x=c_0(1+kat_0)(t-t_0), x=c_0(t-t_1)$  and  the  shock curve $C_S$, where $t_1$ is  the  systole duration, $x=0$ is  the  position of the aortic root, $c_0$ is  the  pulse wave  velocity in the linear  theory. The shock wave
lies  on the  curve $C_s$. The solutions are given by formulas (\ref{sol_a1})-(\ref{sol_p}) and remain unperturbed outside the domains $1$ and $2$.}
\end{figure}
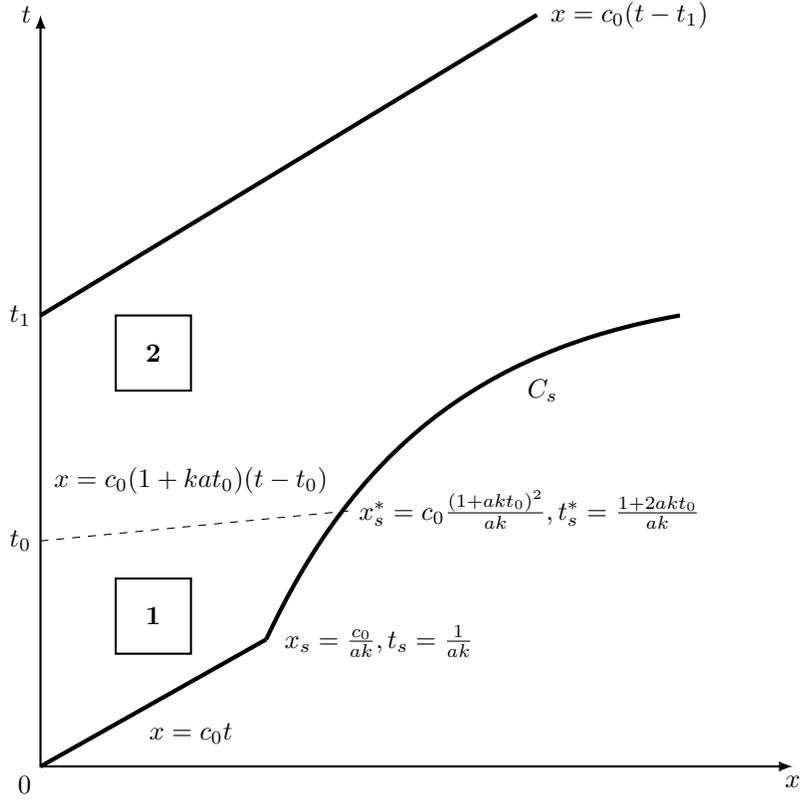

\begin{figure}
\begin{tikzpicture}
\begin{axis}[
width=7cm,height=7cm,
xlabel={$t, [s]$},
xtick={0, 0.15, 0.3},
xmin=0,
ylabel={$p,\, [kPa]$},
ymin=0, ymax=5,
minor y tick num=1,
]

\addplot [blue,mark=*] table {evol_data1.txt};
\end{axis}
\node [left] at (5,5) {$x=0\, m$};

\end{tikzpicture}
\begin{tikzpicture}
\begin{axis}[
width=7cm,height=7cm,
yticklabel pos=upper,
xlabel={$t,\, [s]$},
xmin=0.35, xmax=0.65,
xtick={0.35, 0.5, 0.65},
ymin=0, ymax=5,
minor y tick num=1,
]
\addplot [blue,mark=*] table {evol_data2.txt};
\end{axis}
\node [left] at (5,5) {$x=1.4\, m$};
\end{tikzpicture}
\caption{ The initial symmetrical triangle shaped pulse-wave form at the aortic root (left picture)  at the aortic root $x=0 m$ significantly  changes its form after the wave travels for the time  period of $0.35 s$ and  is  measured at the point $x=1.4 m$ (right picture). The initial pulse wave has the physiological duration of $0.3 s$, $c_0=4 m/s$ and $n=0.5$, the unperturbed radius  of the vessel equals $1.5*10^{-2} m$, $A_0\approx 7*10^{-4}  m^2$. The non-symmetrical initial waves can be   considered using the formulas
(\ref{07sol1})-(\ref{07sol2}).}
\end{figure}
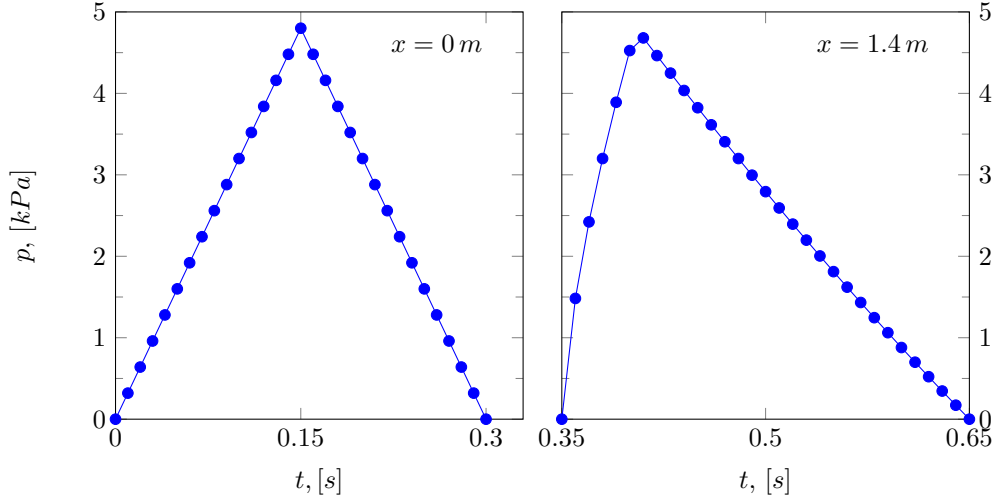

\begin{figure}
\begin{tikzpicture}
\begin{axis}[
width=7cm,height=7cm,
xlabel={$x, [m]$},
xtick={-0.6, 0, 0.6},
xmin=-0.65, xmax=0.65,
ylabel={$p,\, [kPa]$},
ymin=0, ymax=5,
minor y tick num=1,
]

\addplot [blue,mark=*] table {evol_data3.txt};
\end{axis}
\node [left] at (2,5) {$t=0.15 \, s$};

\end{tikzpicture}
\begin{tikzpicture}
\begin{axis}[
width=7cm,height=7cm,
yticklabel pos=upper,
xlabel={$x,\, [m]$},
xmin=0.2, xmax=1.45,
xtick={0.2, 0.8, 1.4},
ymin=0, ymax=5,
minor y tick num=1,
]
\addplot [blue,mark=*] table {evol_data4.txt};
\end{axis}
\node [left] at (2,5) {$t=0.35\, s$};
\end{tikzpicture}
\caption{ Two snapshots  in time for the pressure pulse waves are presented. The aortic outlet is placed at $x=0$. The pressure wave in the left slide is  formally continued for $x<0$.
The initial pulse wave has the physiological duration of $0.3 s$ and $c_0=4 m/s$, $n=0.5$, the unperturbed radius  of the vessel equals $1.5*10^{-2} m$, $A_0\approx 7*10^{-4}  m^2$. The non-symmetrical initial waves can be  considered using the formulas
(\ref{07sol1})-(\ref{07sol2}).}
\end{figure}
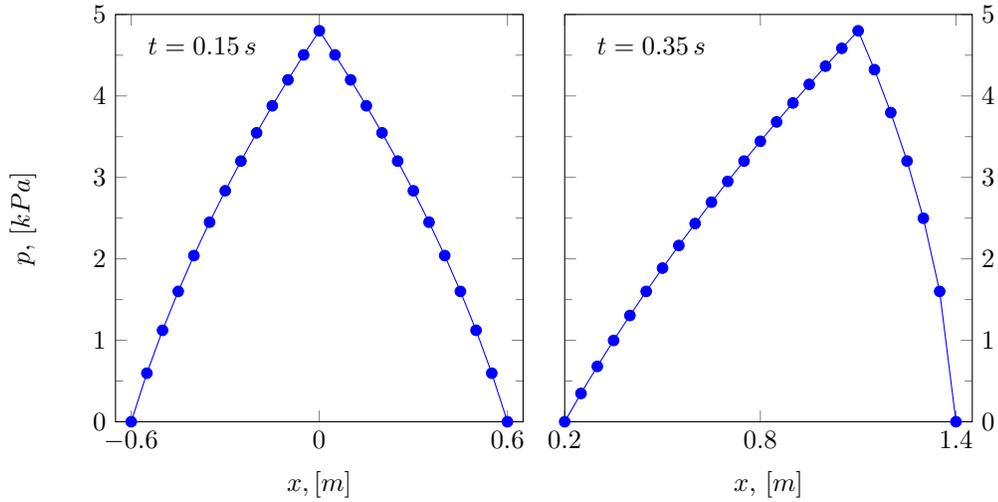

The  typical evolution of the waveform is presented in Fig. 2. and Fig. 3.
Finally, compiling  all the  previous  results  we  obtain the following  proposition

{\it The  problem (\ref{05beq})-(\ref{05bounds4}) has  the  following
solution}
\begin{equation}\label{sol_a1}
A(t,x)= A_0+\frac{-(1-akt)+\sqrt{(1-akt)^2-4ak(c_0^{-1}x-t)}}{2k},
\end{equation}
{\it which is valid for $t \geq 0, x \geq 0$ in the domain bounded by the characteristics $x=c_0t, x=c_0(1+kat_0)(t-t_0)$  and  the part of the shock curve $C_s$ given   by the expression (\ref{05beq_cs}) (the domain $1$ in Figure 1)}

{\it  and}
$$
A= A_0+
$$
\begin{equation}\label{sol_a2}
+\frac{-(1-bk(t_1-t))+\sqrt{(1-bk(t_1-t))^2+4bk(c_0^{-1}x+(t_1-t))}}{2k},
\end{equation}
{\it which is valid for $t \geq 0, x \geq 0$ in the domain bounded by the characteristics $x=c_0(1+kat_0)(t-t_0), x=c_0(t-t_1)$  and  the part of the shock curve $C_s$  given by the equation (\ref{05beq_dom2vel})  in  the case of the symmetrical initial pulse (the domain $2$ in Figure 1).}
{\it We have  used  the following definitions}
$$
k\equiv\frac{n+2}{2A_0}, \quad c_0\equiv \sqrt{A_0^{n}/\rho D_0}.
$$
{\it  Outside of the domains $1$ and $2$  the solution remains  unperturbed, $A=A_0$.}
{\it The blood  velocity $u$ is  calculated  from the linearized version of the  formula (\ref{06subst}) }
\begin{equation}\label{sol_u}
u(t,x)= c_0\frac{A-A_0}{A_0}
\end{equation}
{\it and}
\begin{equation}\label{sol_p}
p(t,x)=p_0+\rho c_0^2\frac{A-A_0}{A_0}.
\end{equation}
{\it This  solutions have  the  error of order $O\bigl((\Delta A/A)^2\bigr)$ for $n \neq 2$ and are  exact for $n=2$.
 The backward wave can be obtained from the formulas above by replacing $c_0^{-1}x$ with $-c_0^{-1}x$ }.

Finally, let us  mention one  important  fact. Since all the terms in  the equations  (\ref{04onedeq2}) (or (\ref{05onedeq}))  depend on $t,x$  only via  $p,A$ (or $u,A$) then the solutions (\ref{sol_a1})-(\ref{sol_u}) are invariant on  shifts
\begin{equation}\label{sol_shift}
(t,x)\rightarrow  (t+t_0, x+x_0).
\end{equation}
The  values  of $t_0,x_0$ are arbitrary constants. For the  practical needs they can be used for fitting the experimental waveforms.
Using  the symmetry  (\ref{sol_shift}) the solutions with  non-triangular wave  forms  at the  aortic root can be easily obtained from the  solutions (\ref{sol_a1})-(\ref{sol_p}) simply by replacing $t,x$  with $t+t_0, x+x_0$.

\section{Exact Nonlinear Wave-Separation Formulas}

We will derive  the closed analytical expression  for the  separation  of a pressure wave measured in an artery into a forward and  backward (reflected wave) components using the Riemann invariants.  Our approach will be similar to the one presented in \cite{1996pythoud} but the final formulas presented in this paragraph have not encountered in literature.

The equations  (\ref{05onedeq}) can be rewritten in the following  form
$$
D_{+}R_{+}=0, \quad  D_{-}R_{-}=0,
$$
where
$$
D_{+}=\frac{\partial}{\partial t}+(u+\frac{1}{\sqrt{\rho D(A)}})\frac{\partial}{\partial x},  \quad D_{-}=\frac{\partial}{\partial t}+ (u-\frac{1}{\sqrt{\rho D(A)}})\frac{\partial}{\partial x},
$$
$$
R_{+}=u+\int_{A_0}^{A}\frac{dz}{\sqrt{\rho D(z)}z}, \quad  R_{-}=u-\int_{A_0}^{A}\frac{dz}{\sqrt{\rho D(z)}z}.
$$
After integration we obtain
\begin{equation}\label{p3exp_riemann}
R_{+}=u+\frac{2}{n\sqrt{\rho D_0}}(A^{n/2}-A_0^{n/2}), \quad R_{-}= u-\frac{2}{n\sqrt{\rho D_0}}(A^{n/2}-A_0^{n/2}).
\end{equation}

We consider  only subsonic flows $|u|<<\frac{1}{\sqrt{\rho D(A)}}$ then  $R_{+}$ is  forward traveling component while $R_{-}$
is  traveling  in the opposite  direction.
From (\ref{p3exp_riemann}) we see  that any velocity  wave  is  composed  from the two components $R_{+},R_{-}$, more exactly  $u=\frac{1}{2}(R_{+}+R_{-})$.
For the case  of the sole forward  wave we need to remove  the backward component, we set $R_{-}=0$, then
$$
R_{+}=2u, \quad  R_{-}=0,
$$
$$
u=\int_{A_0}^{A}\frac{dz}{\sqrt{\rho D(z)}z}
$$
and for the backward wave
$$
R_{+}=0, \quad R_{-}=2u,
$$
$$
 u=-\int_{A_0}^{A}\frac{dz}{\sqrt{\rho D(z)}z}.
$$


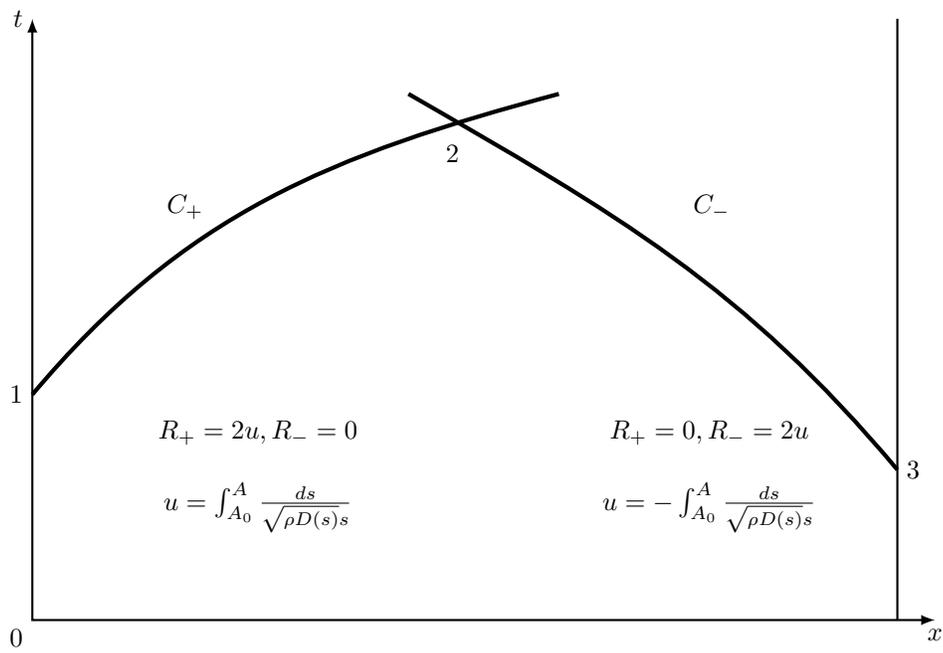
\begin{figure}
\begin{tikzpicture}[>=latex]

 \draw [<->] [thick] (0,8) -- (0,0) -- (12,0);
 \node [below] at (12,0) {$x$};
 \node [left] at (0,8) {$t$};
 \node [below left] at (0,0) {$0$};

 \draw  [thick]  (11.5,0) -- (11.5,8);

 \draw [ultra thick] (0,3) to [out=50,in=195] (7, 7);
 \node [left] at (0, 3) {$1$};
 \node [left] at (2.4, 5.5) {$C_+$};
 \node  at (3, 2.5) {$R_{+}=2u, R_{-}=0$};
 \node at (3, 1.5) {$u=\int_{A_0}^{A}\frac{ds}{\sqrt{\rho D(s)}s}$};

 \draw [ultra thick] (11.5,2) to [out=130,in=330] (5, 7);
 \node [right] at (11.5, 2) {$3$};
 \node [left] at (9.4, 5.5) {$C_-$};
 \node  at (9, 2.5) {$R_{+}=0, R_{-}=2u$};
 \node at (9, 1.5) {$u=-\int_{A_0}^{A}\frac{ds}{\sqrt{\rho D(s)}s}$};

 \node [left] at (5.8, 6.2) {$2$};
\end{tikzpicture}
\caption{ The forward and backward waves starting from the points $1$ and  $3$  travel along
the corresponding  characteristic curves $C_{+},C_{-}$ and  meet  at  the point $2$.}
\end{figure}

Finally, integrating  the  right-hand  sides  of the  expressions for the velocities we  obtain for the forward and  the backward waves
respectively
\begin{equation}\label{p3exp_u}
u=\pm \frac{2}{n\sqrt{\rho D_0}}(A^{n/2}-A_0^{n/2}).
\end{equation}
Let us remember  that the pressure    can be  expressed via vessel cross-section area $A$
using the relation (\ref{p3exp_p})
$$
p=p_0+\frac{1}{n D_0}(A^{n}-A_0^{n}).
$$
Now  we consider  the wave at the point $2$ composed of the forward and backward waves starting  from the points $1$ and $3$ (Fig. 3).
Let us  denote  the total blood  pressure,  the  blood  velocity and the vessel cross-section area  of the  composed  wave at  the  point $2$  as $p,u, A$.
Our  goal is  to separate this wave  into the forward and the  backward  components, or to express the blood pressure of the forward wave $p_+$  and the backward wave $p_{-}$ in terms  of  $p,u$. Using  (\ref{p3exp_p}) we  can  express the Riemann invariants in (\ref{p3exp_riemann}) in terms  of $u,p$
\begin{equation}\label{p3exp_sol_riemann}
R_{+}(u,p)=u+\frac{2}{n\sqrt{\rho D_0}}h(p)  , \quad R_{-}(u,p)=u-\frac{2}{n\sqrt{\rho D_0}}h(p),
\end{equation}
where $h(p)=\sqrt{nD_0(p-p_0)+A_0^n}-A_0^{n/2}$.
Moreover we can invert  this  relations and express $u,p$  in terms  of $R_{+},R_{-}$
\begin{equation}\label{p3exp_sol_p}
p(R_{+},R_{-})=p_0+\frac{1}{nD_0}\left\{ (\frac{n\sqrt{\rho D_0}}{4} (R_{+}-R_{-})+A_0^{n/2})^2-A_0^n  \right\},
\end{equation}
\begin{equation}\label{p3exp_sol_u}
u(R_{+},R_{-})=\frac{R_{+}+R_{-}}{2}.
\end{equation}
Now  let us remember  that the  forward traveling  wave  is  defined  by the condition $R_{-}=0$  while the backward wave is  defined  by $R_{+}=0$. Then we deduce that
\begin{equation}\label{p3exp_sol_psep1}
p_{+}=p(R_{+},0)=p_0+\frac{1}{nD_0}\left\{ \bigl(\frac{n\sqrt{\rho D_0}}{4} R_{+}+A_0^{n/2}\bigr)^2-A_0^n  \right\},
\end{equation}
\begin{equation}\label{p3exp_sol_psep2}
p_{-}=p(0,R_{-})=p_0+\frac{1}{nD_0}\left\{ \bigl(-\frac{n\sqrt{\rho D_0}}{4} R_{-}+A_0^{n/2}\bigr)^2-A_0^n  \right\}.
\end{equation}
We substitute the relations (\ref{p3exp_sol_riemann}) for $R_{+}=R_{+}(u,p), R_{-}=R_{-}(u,p)$
in (\ref{p3exp_sol_psep1})-(\ref{p3exp_sol_psep2})  and after some  algebra we obtain the full nonlinear wave-separation formulas
\begin{equation}\label{p3exp_sol_fin1}
\Delta p_{+}(u,p)=\frac{\rho c_0^2}{2}\left \{\frac{2}{n}g(\Delta p)+\frac{u}{c_0} \right\}\left \{1+\frac{1}{4}g(\Delta p)+\frac{nu}{8c_0} \right \},
\end{equation}
\begin{equation}\label{p3exp_sol_fin2}
\Delta p_{-}(u,p)=\frac{\rho c_0^2}{2}\left \{\frac{2}{n}g(\Delta p)-\frac{u}{c_0} \right\}\left \{1+\frac{1}{4}g(\Delta p)-\frac{nu}{8c_0} \right \},
\end{equation}
where
$$ g(\Delta p)=\sqrt{\frac{n\Delta p}{\rho c_0^2}+1}-1$$
and $\Delta p=p-p_0, \Delta p_{\pm}=p_{\pm}-p_0$. If  we expand $g(\Delta p)$ in Taylor series on $\Delta p$  and  keep only  the first order term then  $g(\Delta p) \approx \frac{n\Delta p}{2c_0^2}$. Therefore the linear approximation of (\ref{p3exp_sol_fin1})-(\ref{p3exp_sol_fin2}) gives the classical linear wave-separation formulas \cite{1972westerhof}
$$
\Delta p_{+}(u,p)=\frac{\rho c_0^2}{2}  \Bigl(\frac{\Delta p}{\rho c_0^2}+\frac{u}{c_0}\Bigr),\quad  \Delta p_{-}(u,p)=\frac{\rho c_0^2}{2} \Bigl(\frac{\Delta p}{\rho c_0^2}-\frac{u}{c_0}\Bigr).
$$

Keeping the  quadratic terms  in the expressions (\ref{p3exp_sol_fin1})-(\ref{p3exp_sol_fin2}) leads to the following simplified  nonlinear wave separation formulas
\begin{equation}\label{p3exp_sol_fin_quad1}
\Delta p_{+}(u,p)=\frac{\rho c_0^2}{2} \left \{ \Bigl(\frac{\Delta p}{\rho c_0^2}+\frac{u}{c_0}\Bigr)
+\frac{n}{8}\Bigl [-\Bigl(\frac{\Delta p}{\rho c_0^2} \Bigr )^2+ 2\frac{u}{c_0}\frac{\Delta p}{\rho c_0^2}+\Bigl(\frac{u}{c_0}\Bigr)^2\Bigr]\right \},
\end{equation}
\begin{equation}\label{p3exp_sol_fin_quad2}
\Delta p_{-}(u,p)=\frac{\rho c_0^2}{2} \left \{ \Bigl(\frac{\Delta p}{\rho c_0^2}-\frac{u}{c_0}\Bigr)
+\frac{n}{8}\Bigl [-\Bigl(\frac{\Delta p}{\rho c_0^2} \Bigr )^2- 2\frac{u}{c_0}\frac{\Delta p}{\rho c_0^2}+\Bigl(\frac{u}{c_0}\Bigr)^2\Bigr]\right \}.
\end{equation}
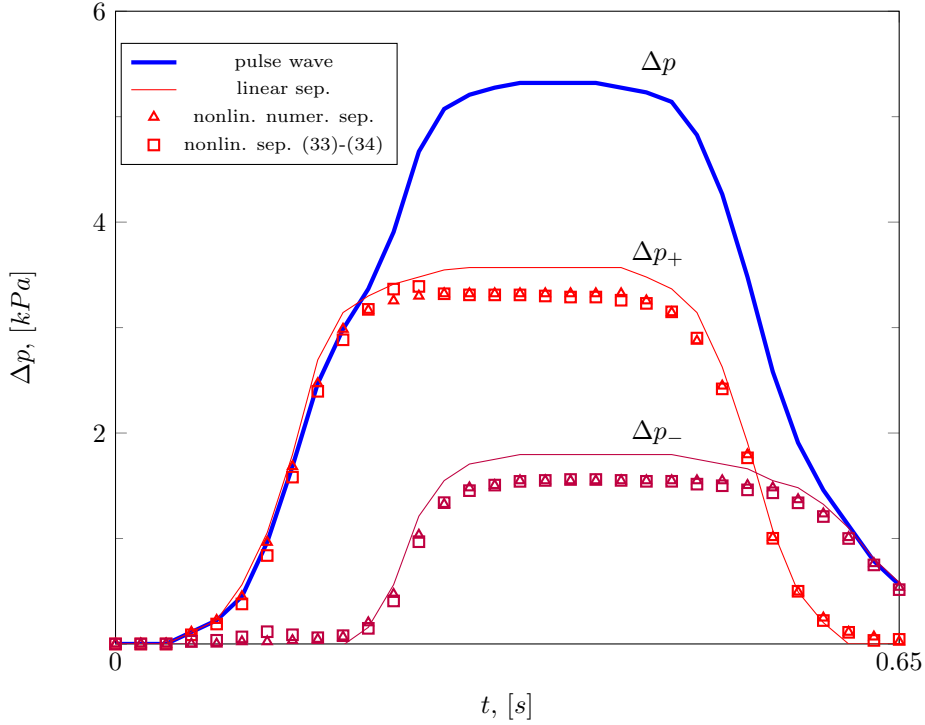
\begin{figure}
\begin{tikzpicture}
\begin{axis}[
legend style={at={(0.36, 0.95)} },
width=12cm,height=10cm,
xlabel={$t,\, [s]$},
xtick={0, 0.65},
xmin=0, xmax=0.65,
ylabel={$\Delta p,\, [kPa]$},
ymin=0, ymax=6000,
ytick={0,2000,4000,6000},
yticklabels={,$2$,$4$,$6$},
minor y tick num=1
]
\addplot [blue, ultra thick] table {sep_tot.txt};

\addplot [red] table {sep_plus_lin.txt};
\addplot [only marks, thick, red,mark=triangle] table {sep_plus_num.txt};
\addplot [only marks, thick, red,mark=square] table {sep_plus_closed.txt};

\addplot [purple] table {sep_min_lin.txt};
\addplot [only marks, thick, purple,mark=triangle] table {sep_min_num.txt};
\addplot [only marks, thick, purple,mark=square] table {sep_min_closed.txt};


\addlegendentry{{\scriptsize pulse wave}};
\addlegendentry{{\scriptsize linear sep.}};
\addlegendentry{{\scriptsize nonlin. numer. sep.}};
\addlegendentry{{\scriptsize nonlin. sep. (\ref{p3exp_sol_fin1})-(\ref{p3exp_sol_fin2})}};

\node at (axis cs: 0.45, 5500)  {$\Delta p$};
\node at (axis cs: 0.45, 3700)  {$\Delta p_+$};
\node at (axis cs: 0.45, 2000)  {$\Delta p_-$};

\end{axis}
\end{tikzpicture}
\caption{ The pressure  pulse wave (thick upper line, borrowed from \cite{2012mynard}) and  its  separation in the forward and the backward components for $n=4, c_0=4 m/s, \rho=1060 kg/m^3$.
The squared symbols correspond  to the results  obtained using the nonlinear wave separation formulas (\ref{p3exp_sol_fin1})-(\ref{p3exp_sol_fin2}), the other waves are  obtained  using  the classic  linear method  and  the non-linear numerical approach (lines without markers and  the triangle symbols respectively) from \cite{2012mynard}. The both nonlinear methods show very close  results.}
\end{figure}

The  results of the  analytical study can be  validated on the data  from  literature.
We compare  the  wave profiles calculated  using the  formulas (\ref{p3exp_sol_fin1})-(\ref{p3exp_sol_fin2})
with the results  presented  in \cite{2012mynard}  where  the  nonlinear wave separation technique was based
on the calculation of the infinitesimal changes  of the forward and the backward waves
$dp_{\pm}=\pm\frac{\rho c_{\pm}}{2}dR_{\pm}$, where $c_{\pm}=u \pm c(p),  c(p)=\sqrt{(n/\rho)\Delta p+ c_0^2}$.
The equations for $R_{\pm}$ : $dR_{\pm}/dt+c_{\pm} dR_{\pm}/dx=0$ are solved  numerically using  the  locally conservative Galerkin
method. The details about  the applied numerical method can be also found  in \cite{2008mynard},\cite{2010mynard}.
The numerical approach from  \cite{2012mynard} and the present approach give  very close results  since  the same equations are solved. The comparison of the methods shows  that the  discrepancies are negligible except of some  points (Fig. 4).

The form of the simplified nonlinear wave separation expressions  (\ref{p3exp_sol_fin_quad1})-(\ref{p3exp_sol_fin_quad2}) allows to assess the influence of the  nonlinear effects. Typically $\Delta  p > \rho c_0 u$, therefore the  term $-(n/8)(\Delta p/\rho c_0^2)^2$  is  dominant among  the nonlinear terms. Obviously, this term suppresses the  amplitudes of  forward and  backward waves.
This feature  is  illustrated in Fig 4.
The  effect can be observed in several cases. Firstly, for high pressures the  vessel's wall compliance becomes low and this effect can be  modeled using $n>>1$ and  then the nonlinear effects should be taken in account if $c_0$ remains average for the such pressures. On the  other hand, for the classical Hooke  law $n=1/2$ the nonlinear effects can be significant for moderate velocities $c_0=\frac{\sqrt{A}}{\rho D_0}=\frac{2\sqrt{\pi}Eh}{3\rho A_0}$   which can be  potentially  observed in a situation  if the  Young's modulus $E$ and the wall thickness $h$ are small and the vessel is compliant.

\section*{Conclusion}
The nonlinear 1D  equations for the inviscid blood flow were investigated analytically.
The viscoelastic terms in the vessel wall-pressure  response were excluded in the analysis,  the  unperturbed luminal area was constant alongside  the considered vessels.
Closed expressions for  nonlinear forward (or backward) waves for a class  of  particular initial-boundary conditions (including  triangle shape profiles)  has been derived assuming that the terms $(\Delta A/A_0)^2$ (the quadratic terms on the relative systolic  change of  the vessel's cross-section area) are small. As a result, the shock formation conditions have been obtained. In addition, the analytical expressions for the wave separation into
the  forward and  backward components were deduced.

The  introduction of friction in the simplest form results in the addition of  $\frac{ frict}{\rho A}$  terms  in the right side  of the second equation (\ref{05onedeq}) ,  where $frict$ is the wall friction force proportional to $ -\nu u$ and $\nu$ is the blood viscosity. For the Eqs (\ref{05onedeq}) with the friction included the forward traveling wave cannot be taken in the form (\ref{06subst}). Qualitatively we  suppose  that the friction will prevent the appearance
of shocks if the initial wave is too flat \cite{1999whitham}.
Mathematically this problem can be considered as the asymptotic expansion on $\nu$  which is an interesting subject the future study.

The analytical  nonlinear pulse wave expression could be used in the evaluation of the central pressure based on the pressure
forms in other arteries.
The  principal method for the assessment  of  the central (aortic)  blood  pressure from the applanation tonometry measurements
in brachial (or radial) arteries is  based on the application of transfer functions \cite{1993Karamanoglu}-\cite{2001pauca}. This function matches the Fourier harmonics of the radial (brachial) and the aortic pressures
and is derived using  the  regression analysis. As  the  shape of the nonlinear pressure wave and  its change  during the propagation can be  obtained  analytically then the following approach could be tested. As a first step the separation of the  blood pressure measured in a radial artery into  forward and  backward components could be  performed using  the pressure profiles only \cite{2006westerhof}.  Next, we estimate  the free parameters $t_0, x_0$  (see (\ref{sol_shift})) for which the analytical profile has the best fit with the assessed forward component. Finally, the aortic pressure can be  recovered  by application of the  shift
in variables $(t,x) \rightarrow (t-\delta t, x-\delta x)$  where $(\delta t, \delta x)$ are the approximate time and distance  traveled by the pulse wave from the aortic root to the vessel under the consideration.

I wish to thank Oleg A. Rogozin for the  discussions.

The work was supported by Russian Foundation for Basic Research, grant No 18-01-00899.

\section*{Appendix. Distensibility and Hooke law}

Consider the  pressure-area relationship (\ref{p3exp_p})
$$
p=p_0+\frac{1}{n D_0}(A^{n}-A_0^{n}).
$$

 If $n=1/2$ the  Hooke  law  is derived in the following  form \cite{2011alastruey}
\begin{equation}\label{p3_hooke}
p=p_0+\frac{4\sqrt{\pi}Eh}{3A_0}(\sqrt{A}-\sqrt{A_0}),
\end{equation}
where $E$ is the Young's modulus, $h$ is the vessel wall thickness. The  viscoelastic terms are  not considered  in the present study.
In general case  the Young's modulus  is  not constant but is an increasing function of   blood  pressure $E=E(p)$ \cite{1984lang}. This  fact means that the stiffness of an artery increases for high pressures. In contrast to the Hooke  law $A(p) \sim p^2$ , for the real arteries like aorta (non-constant Young's modulus) the saturation  of  vessel distension is  observed, this effect  can be  modeled using the arctangent  function $A(p) \sim \arctan(p)$. The saturation effects appear for  pressures close to $200 \, mmHg$, , for  pressures in a range  $100-200 \, mmHg$  we observe  that $\frac{dA}{dp}$  is decreasing function of $p$, see Fig. 4  in \cite{1984lang}. This feature can be recovered in the expressions (\ref{p3exp_p}) if we assume that $n>1$  ($A \sim p^{1/n}$).

\section*{Appendix 2. General solution for the sole forward wave}

Recall the equation (\ref{05onedeq})
$$
\pder[A]{t}+\frac{1}{\sqrt{\rho D_0}}\left (\left(1+\frac{2}{n}\right)A^{n/2}-\frac{2}{n}A_0^{n/2}  \right )\pder[A]{x}=0,
$$
the formal solution along  the characteristics is
$$
A=const, \quad \frac{dt}{dx}=\frac{n\sqrt{\rho D_0}}{(n+2)A^{n/2}-2A_0^{n/2}},
$$
where $dt/dx>0$  since $A \geq A_0$ for any values of $t,x$.
Note that $A$ is  constant  on the  characteristics $t(x)$. The characteristics  depend on the different  values of the parameter $s$, $t(0)=s$.
Then, after  integration  we  have
\begin{equation}\label{05formal_sol}
A=f(s),  \quad  t=\frac{n\sqrt{\rho D_0}}{(n+2)f(s)^{n/2}-2A_0^{n/2}}x+s,
\end{equation}
where  the function $f(s)$  should  be defined  from the  boundary conditions.
Replacing the parameter $s$ by $f^{-1}(A)$ in the second equation  (\ref{05formal_sol}) we obtain
$$
A=f\left(t-\frac{n\sqrt{\rho D_0}x}{(n/2)A^{n/2}-2A_0^{n/2}}  \right).
$$
For the boundary  conditions  (\ref{05bounds1})
 we conclude  that $f(s)$ takes the form
$$
f(s)=A_0+as,
$$
then
\begin{equation}\label{05formal_sol2}
A=A_0+a\left(t-\frac{n\sqrt{\rho D_0}x}{(n+2)A^{n/2}-2A_0^{n/2}} \right).
\end{equation}
The explicit analytical expression for $A$ can not be obtained from the equation (\ref{05formal_sol2}) in the case  of  general $n$.

\end{document}